\documentclass[twoside,slac_one]{revtex4}
\usepackage{graphicx}
\usepackage{fancyhdr}
\usepackage{amsmath} % American Mathematics Society standards
\usepackage{bm}% bold math
\usepackage{amsxtra}
\usepackage{amssymb}
\usepackage{amsthm}
\usepackage{latexsym}
\usepackage{lscape}

\begin{document}

\title{Monte-Carlo simulation of jets in heavy-ion collisions}


\author{Clint Young}
\affiliation{Department of Physics, McGill University, 3600 University Street, Montreal, Quebec, H3A\,2T8, Canada}
\author{Bj\"orn Schenke\footnote{presenter}}
\affiliation{Physics Department, Building 510A, Brookhaven National Laboratory Upton, NY 11973, USA}
\author{Sangyong Jeon}
\author{Charles Gale}
\affiliation{Department of Physics, McGill University, 3600 University Street, Montreal, Quebec, H3A\,2T8, Canada}

\begin{abstract}
We present Monte-Carlo simulations of jet evolution in lead-lead collisions at the Large Hadron Collider (LHC) at CERN focusing on the 
dijet asymmetry measured by the ATLAS and CMS collaborations.
In the simulation, hard partons are interacting with the hydrodynamical background medium, undergoing radiative and collisional processes. The measured dijet asymmetry is well described by the simulation and is hence consistent with partonic energy loss in a hot, strongly-interacting medium.
\end{abstract}

\maketitle

\thispagestyle{fancy}


%%%%%%%%%%%%%%%%%%%%%%%%%%%%%%%%%%
\section{Introduction}

With a center of mass energy of $\sqrt{s_{NN}}=2.76\;{\rm TeV}$ at the LHC, which is significantly 
larger than that achieved at the RHIC, far more energetic jets are kinematically accessible at the LHC. 
The ATLAS collaboration was able to measure over 1000 dijets where the leading jet has transverse energy $E_T > 100\;{\rm GeV}$ and the opposing 
jet has energy $E_T > 25\;{\rm GeV}$ \cite{Collaboration:2010bu}. The CMS collaboration performed a similar analysis on a large sample of 
jets ($E_{T1}>120\;{\rm GeV}$, $E_{T2}>50\;{\rm GeV}$) \cite{Chatrchyan:2011sx}. 
These results are a significant improvement over the results from the RHIC, where the total energies of the jets were far lower and therefore harder to separate from fluctuations in the underlying bulk. Also, the models for 
partonic evolution rely on the probe parton having high energy, and when this separation of energies exists one can expect the hadronization of 
these partons to be described well with vacuum fragmentation functions. 

% Several theoretical studies of these results are now available: Majumder and Che et al. examined the supression of high-$p_T$ hadrons assuming purely radiative energy loss and found good agreement with the rising $R_{AA}$ for high transverse momentum seen in the latest analysis of ALICE \cite{Majumder:2011uk, Che:2011vt}.

Several theoretical studies have addressed the measured dijet asymmetry at the LHC. In \cite{Qin:2010mn} 
the evolution of the jet shower was studied, analyzing the jet propagation through the quark-gluon plasma and its interaction with the 
medium. Good agreement with the experimental data was found. The authors of  
\cite{CasalderreySolana:2010eh} conclude that the removal of soft components from within the jet cone via elastic collisions will induce a dijet asymmetry.
Also, the \textsc{pyquen} model \cite{Lokhtin:2011qq} was used the to quantify the ``jet-trimming''.

In this work, we present a somewhat more detailed analysis of what was presented in \cite{Young:2011qx}, where we applied \textsc{martini} to lead-lead collisions at the LHC \cite{Schenke:2009gb, Schenke:2009vr}.
In Section \ref{transport}, the physics behind \textsc{martini} is reviewed, as well as the description of the bulk of 
heavy-ion collisions with 3+1-dimensional hydrodynamics. In Section \ref{results}, runs of \textsc{martini} with cuts given by the ATLAS and 
CMS detectors and their analyses are compared with the experimental results for both $dN/dA_J$ (the yield of dijets differential in $A_J$, 
where $A_J$ measures the energy anisotropy of dijets) and $dN/d\phi$. 

\section{Transport of high-energy partons and \textsc{martini}}
\label{transport}

\textsc{martini} solves the rate equations
\begin{eqnarray*}
\frac{dP_{q\bar{q}}(p)}{dt} &=& \int_k P_{q\bar{q}}(p+k) \frac{d\Gamma^q_{qg}(p+k,k)}{dkdt}-P_{q\bar{q}}(p)\frac{d\Gamma^q_{qg}(p,k)}{dkdt}+P_g(p+k)
\frac{d\Gamma^g_{q\bar{q}}(p+k,k)}{dkdt}{\rm ,} \\
\frac{dP_g(p)}{dt} &=& \int_k P_{q\bar{q}}(p+k)\frac{d\Gamma^q_{qg}(p+k,p)}{dkdt}+P_g(p+k)\frac{d\Gamma^g_{gg}(p+k,k)}{dkdt} \\
 & & -P_g(p)
\left( \frac{d\Gamma^g_{q\bar{q}}(p,k)}{dkdt}+\frac{d\Gamma^g_{gg}(p,k)}{dkdt}\Theta(2k-p) \right){\rm ,} \\
\end{eqnarray*}
where the various {\it differential} rates $d\Gamma^i_{lm}(p,k)/dkdt$ determine the splitting of partons $l$ and $m$, one with momentum $k$, from 
a parton $i$ with momentum $p$ \cite{Turbide:2005fk}.
%\begin{equation}
%\frac{dP(E)}{dt}=\int d\omega \left[ P(E+\omega)\frac{d\Gamma(E+\omega, E)}{d\omega}-P(E)\frac{d\Gamma(E,E+\omega)}{d\omega} \right]
%\end{equation}
%using Monte Carlo methods. The above equation is schematic: in \textsc{martini}, we evolve quarks and gluons separately keeping track of all radiated hard 
%partons. The rates $Pd\Gamma/d\omega$ in the above equation therefore represent the sum of all processes that are included in \textsc{martini}. 

In its current implementation, \textsc{martini} uses rates which take into account both radiative and collisional QCD processes, calculated at finite temperature.
Collisional processes involve soft momentum transfers sensitive to the gluon's screening mass and 
therefore, hard thermal loop results at leading order are used to describe these processes.
For the elastic processes \textsc{martini} does not depend on the ``diffusive approximation'': there is no need to assume that the rates are 
only significant when $\omega$ is small \cite{Schenke:2009ik}.
It is important to point out at this point that for the elastic collisions there is no exact energy and momentum conservation within the hard partons because energy is lost to the strongly coupled hydrodynamic medium. Because at this point this lost energy and momentum is not added to the energy in the hydrodynamic calculation (the hydrodynamic background is not modified by the hard partons), it is lost from the analysis. The underlying assumption here is that this energy and momentum is thermalized rapidly and becomes part of the background medium.
Radiative energy loss is modeled using the Arnold-Moore-Yaffe approach, where the interference of bremsstrahlung gluons from multiple scatterings is taken into account with an LPM-like integral equation for the energy loss rate \cite{Arnold:2001ms, Arnold:2001ba, Arnold:2002ja}.
Also for the radiative processes a certain cut in momentum has to be included, because the AMY formalism is not applicable for parton energies of the order of the temperature. We assume that radiated partons with momenta below $2\,{\rm GeV}$ are better described as part of the strongly coupled medium and do not integrate them into the hard parton shower. Furthermore, partons with energies below $4\,T$ do not undergo any perturbative interactions anymore.
A more detailed analysis of the dependence of final results on these separation scales between the perturbative regime and the strongly coupled bulk medium is currently on the way and will be presented in a forthcoming work.
%It is impotant to point out that low momentum partons get transverse kicks on the same order of magnitude as the high momentum partons, leading to larger changes of angles for the low momentum partons and deflections kicks out of the jet cone.

The momenta of high-energy partons are sampled using \textsc{pythia} event generation \cite{Sjostrand:2006za}, and their initial positions in the transverse plane of heavy-ion collisions are sampled according to $n_B(x, y, b)$, the distribution of binary collisions for a given impact parameter $b$ of the collision. 
These partons are then evolved through the background of bulk particles. In this work, this evolving background is modeled 
using \textsc{music}, a 3+1-dimensional hydrodynamic simulation \cite{Schenke:2010nt}.

For the results in Section \ref{results}, \textsc{martini} is run with $\alpha_s=0.27$ including both collisional and 
radiative processes, as well as with radiative and collisional energy loss separately. 
The finite-temperature rates for these processes are determined by the temperatures and flow in lead-lead collisions as 
simulated with \textsc{music} for an impact parameter of $b=2.31\;{\rm fm}$, reproducing the multiplicities of the 0-10\% centrality class.
In this study we use a simulation with ideal hydrodynamics starting with averaged initial conditions.

% In summary, the strengths of \textsc{martini} include the inclusion of combined radiative and elastic processes, its 
% integration with \textsc{pythia} and Glauber model calculations for both sampling of the initial parton distributions in momentum and 
% position and the fragmentation of the evolved partons into hadrons, and the ability to evolve the partons in a background medium obtained 
% from realistic hydrodynamical simulations.

\section{Results for lead-lead collisions measured at ATLAS and CMS}
\label{results}

Once high-energy partons have evolved and hadronized, the resulting hadrons must then be reconstructed into jets. For the best possible comparison with the results of the LHC, we use the same anti-$k_t$ jet reconstruction that the ATLAS collaboration uses \cite{Cacciari:2008gp}. These algorithms 
depend on the definition of distances between two 4-momenta:
\begin{equation}
d_{ij}={\rm min} \left( \frac{1}{k^2_{it}}, \frac{1}{k^2_{jt}} \right) \frac{(\phi_i-\phi_j)^2+(y_i-y_j)^2}{R^2}{\rm .}
\end{equation}
The distances are determined between all pairs of final-state particles whose energies are large enough to trigger the calorimeters, and starting with the smallest distance, 4-momenta close to each other are clustered and added together and final jets are determined. The implementation of this algorithm that we used is \textsc{fastjet}, publicly available online \cite{fastjet}.

Once the clustering of hadrons into jets is complete, the jet with highest $E_T$ is determined, and the highest energy jet whose azimuthal angle from the leading jet $\Delta \phi > \pi/2$ (or $2\pi/3$, as is the case with the CMS analysis) is also determined. If the energies of this dijet are high enough to make it into the given detector's analysis, they are recorded and binned.
 
\begin{figure}
\includegraphics[height=9cm]{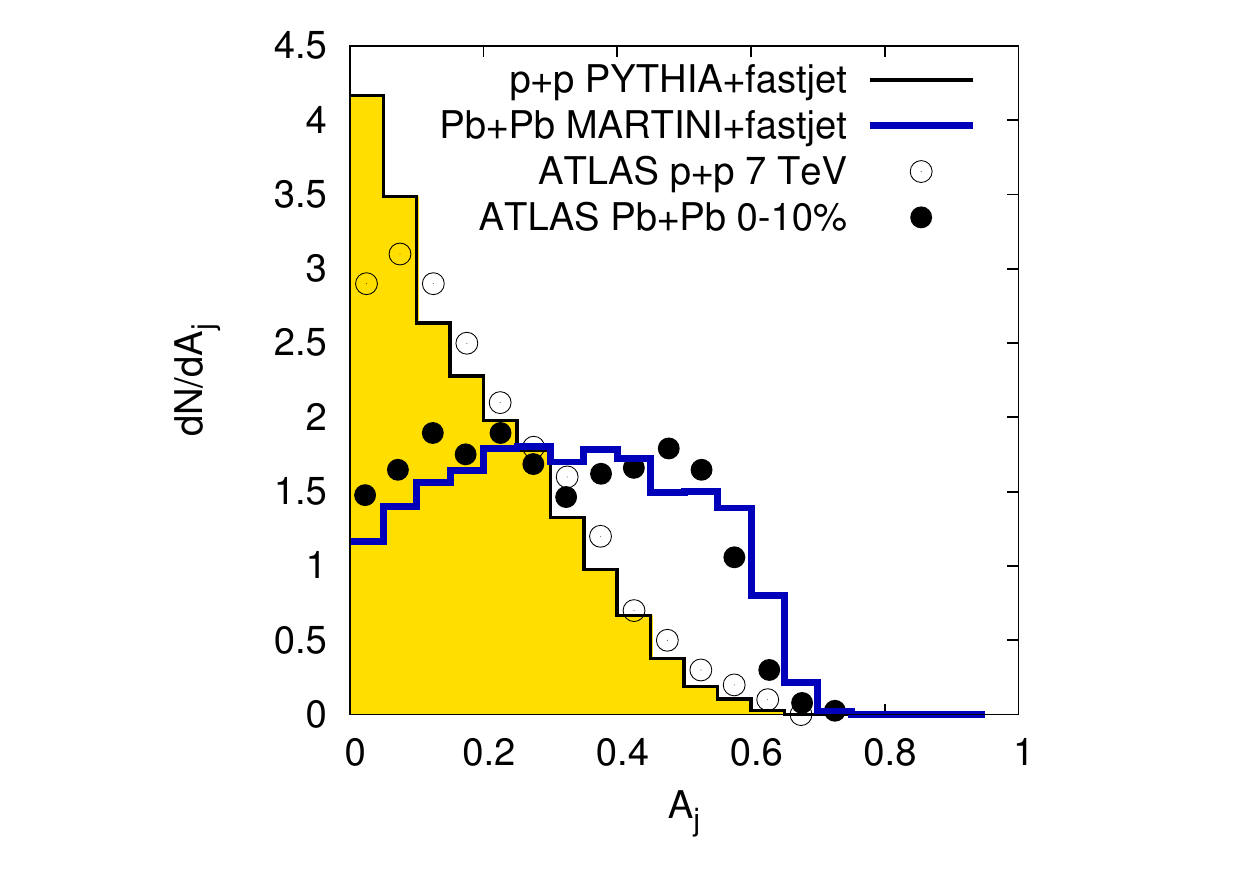}
\caption{
The differential yield $dN/dA_J$ for proton-proton collisions and 
lead-lead collisions.
}
\label{dNdA_ET07}
\end{figure}

\begin{figure}
\includegraphics[height=9cm]{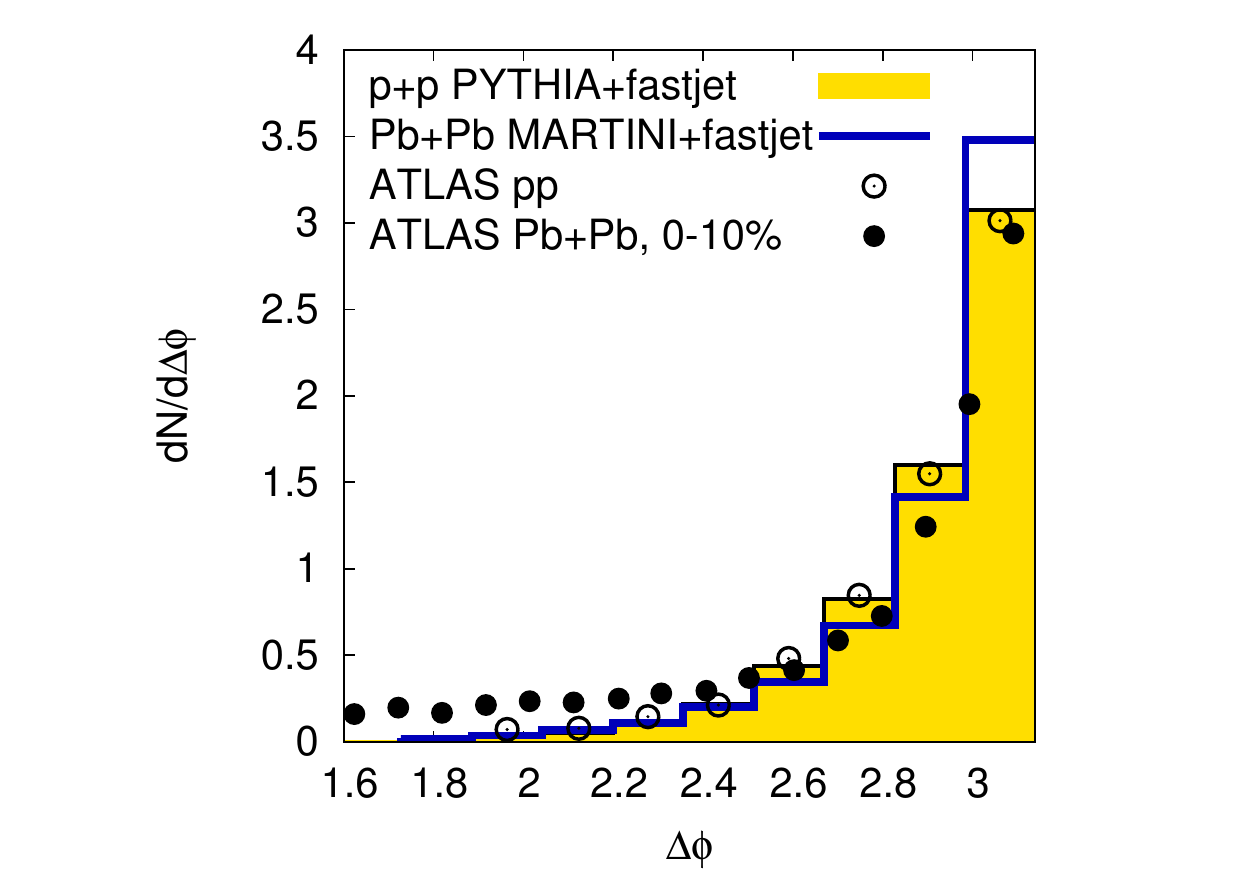}
\caption{
The differential yield $dN/d\Delta\phi$ for proton-proton collisions and lead-lead collisions.
 }
\label{dNdphi_ET07}
\end{figure}

In Figure \ref{dNdA_ET07}, we show the results for ATLAS, in the 0-10\% centrality range, for the differential yield $dN/dA_J$, where $A_J=\frac{E_{T1}-E_{T2}}{E_{T1}+E_{T2}}$ is a measure of the transverse energy asymmetry of the dijets. The ATLAS results used are from the 
latest analysis using $R=0.4$ \cite{Cole:QM2011}; there was little dependence on $R$ found in the latest results, suggesting partonic energy loss as the dominant mechanism leading to dijet asymmetry. Our results are compared with p+p events using \textsc{pythia} and \textsc{fastjet}, and the differential yields are normalized to one. In Figure \ref{dNdphi_ET07}, we show the differential yields $dN/d\phi$, where $\phi$ is the azimuthal opening angle for the dijets.

\begin{figure}
\includegraphics[height=9cm]{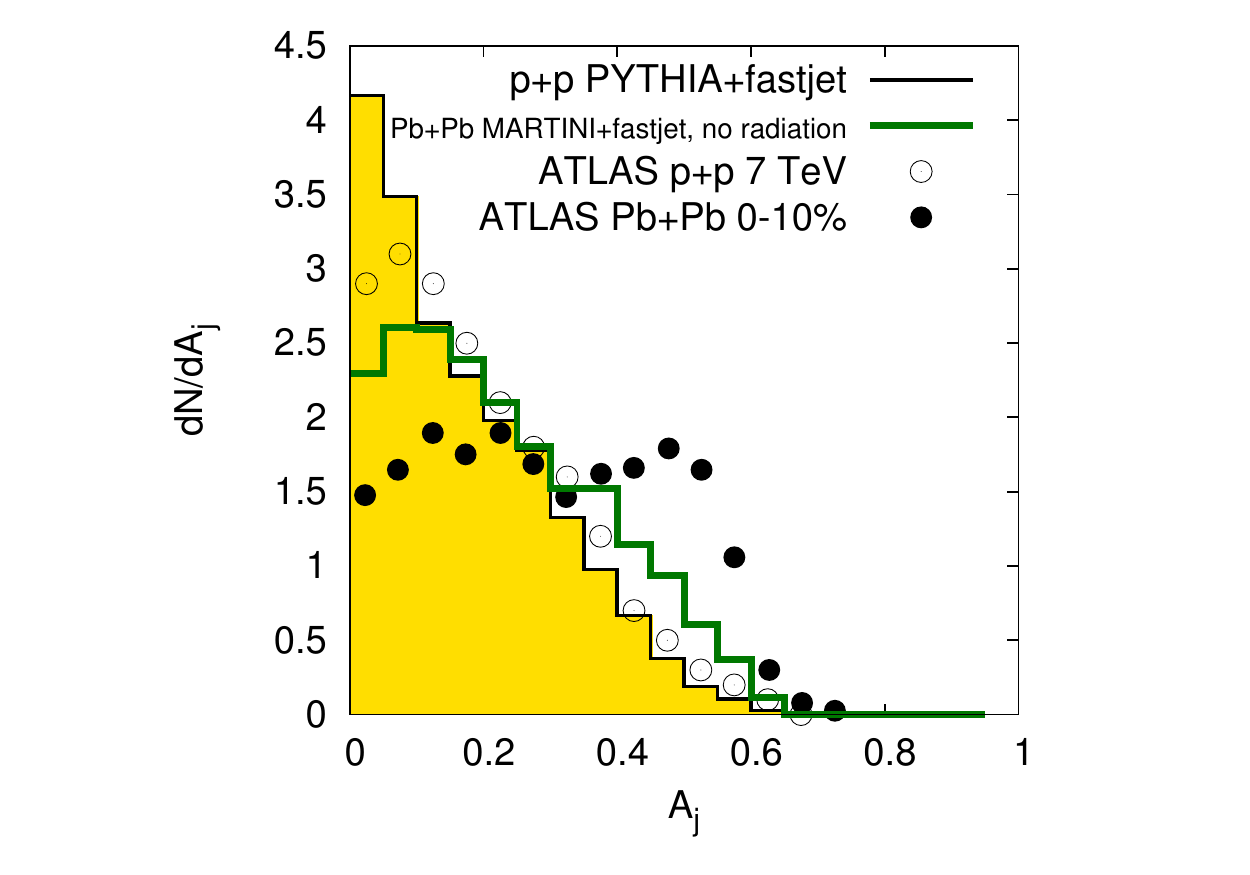}
\caption{
The differential yield $dN/dA_J$ for proton-proton collisions and 
lead-lead collisions, including only elastic processes.
}
\label{norad}
\end{figure}

\begin{figure}
\includegraphics[height=9cm]{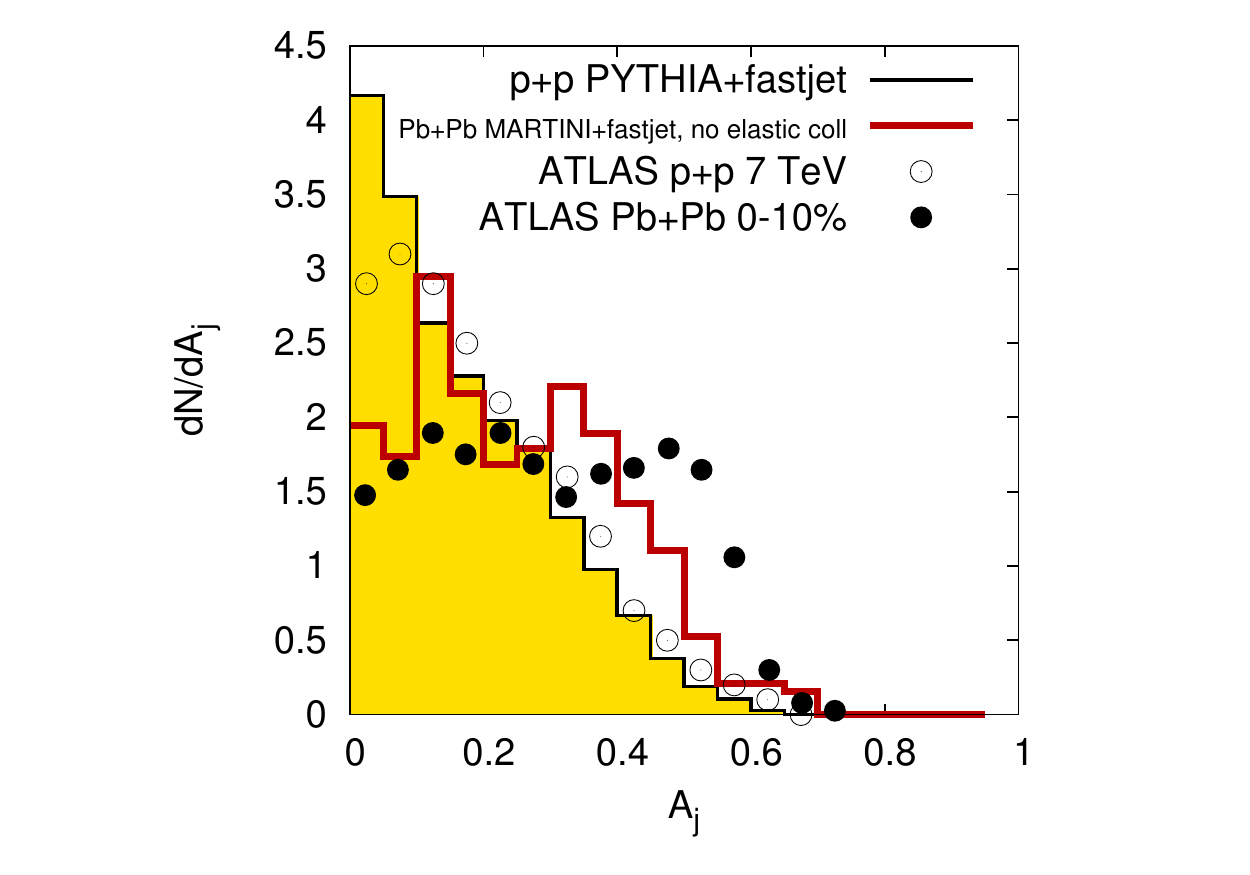}
\caption{
The differential yield $dN/dA_J$ for proton-proton collisions and 
lead-lead collisions, including only radiative processes.
}
\label{noel}
\end{figure}

The presented results are for $\alpha_s=0.27$.
Fig. \ref{dNdphi_ET07} shows no significant difference in the distribution of dijets between proton-proton and lead-lead collisions. The 
experimental results show an increase in the yield at small $\phi$ in lead-lead collisions over what was observed in 
proton-proton collisions. This enhancement, while significant, affects a relatively small number of dijets 
in ATLAS' sample, and can be explained by uncertainties in the combinatorics. For example, if one of the jets is absorbed and an uncorrelated background fluctuation (or part of a second dijet) is used as associated jet instead, this could lead to such enhancement of $dN/d\phi$. Of course it will be most noticeable at $\Delta \phi \sim \pi/2$, where $dN/d\phi$ is smaller.
%  This possible explanation was demonstrated recently by Cacciari, Salam, and Soyez, without any consideration of jet quenching \cite{Cacciari:2011tm}. We are currently working on including the event-by event fluctuations of the initial conditions to take this effect into account. However
% we should point out that these fluctuations affect a relatively small number of jets and does not significantly affect our results besides the differential 
% yield at small angles (which is clear when plotted semi-logarithmically).
To get a feeling for the relevant ingredients needed to reproduce the experimentally observed dijet asymmetry, we present results of calculations with only elastic and only radiative processes in Figs. \ref{norad} and \ref{noel}, respectively. 

The reason why there is a difference to the proton-proton result in both of those cases is the following. Partons can get kicked out of the cone due to elastic collisions and hadrons can get kicked out due to the fragmentation $k_T$ in the first case. There is also some additional asymmetry because the energy of the shower is not completely conserved in the calculation - energy and momentum is lost to the bulk medium and we do not keep track of it, assuming that it thermalizes in the strongly coupled medium.
In the case of only radiative processes, there is broadening due to the fragmentation $k_T$, which can kick hadrons coming from soft radiated partons out of the cone, but also the previously mentioned cutoff below which radiated gluons are not added to the shower causes some asymmetry. A detailed study of the dependence on this cutoff is on the way.

Together, this leads to a description of the process in the full simulation. It is apparent from the figures that both radiative and elastic processes are needed to describe the experimentally observed dijet asymmetry. Radiative processes generate soft partons which can be kicked out of the cone more easily than hard partons by elastic collisions. Furthermore, the fragmentation $k_T$ in \textsc{pythia}'s Lund model will also transport hadrons from softer partons out of the cone more efficiently.

The soft physics of the shower partons is not described in detail within this pQCD based simulation such that some assumptions, manifest in the low momentum cutoffs for certain processes, had to be made. Nevertheless, the general physical process can be well reproduced with this perturbative description of the hard degrees of freedom coupled to a strongly interacting bulk medium.

\begin{figure}
\includegraphics[height=9cm]{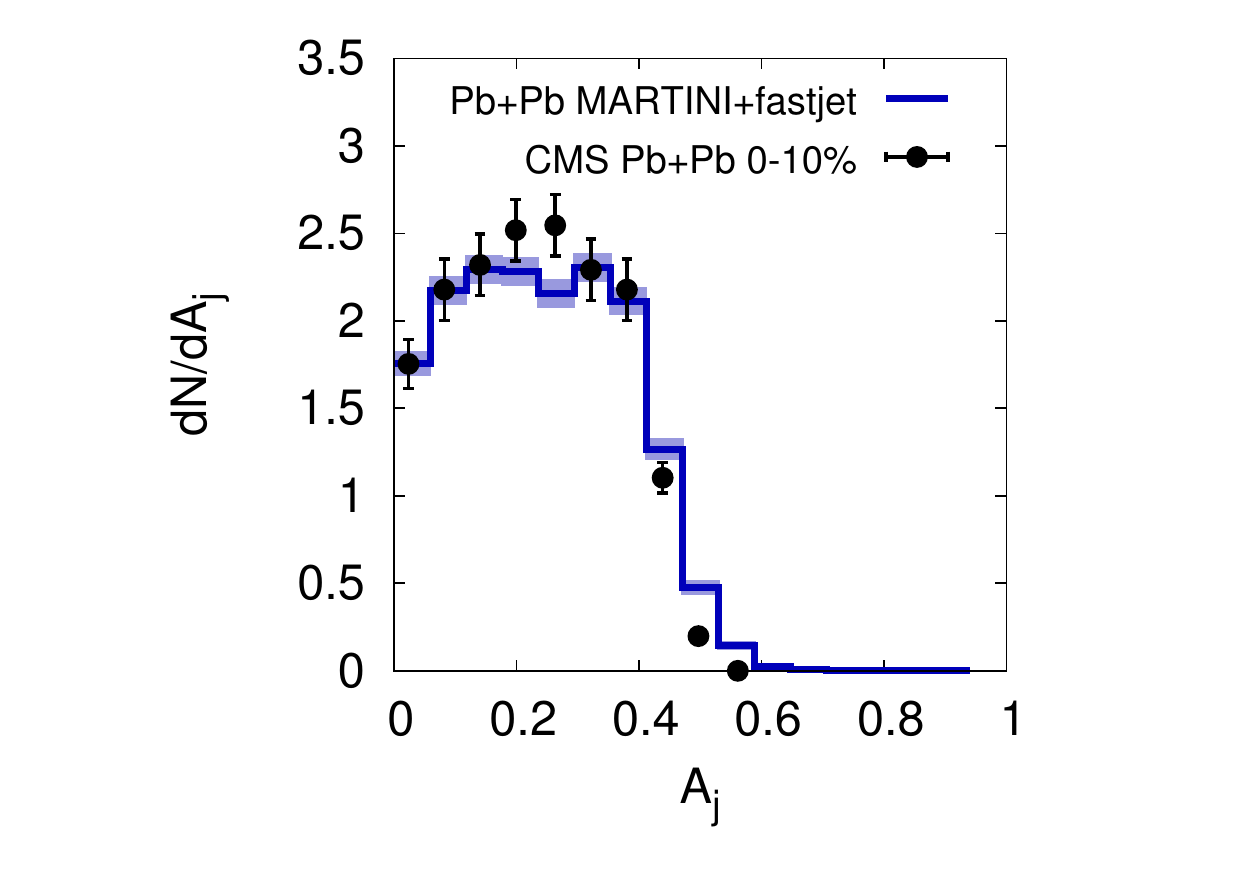}
\caption{
The differential yield $dN/dA_J$ for lead-lead collisions at $\sqrt{s}=2.76\;{\rm GeV}$ compared to CMS data \cite{Chatrchyan:2011sx}.
}
\label{dNdA_CMS}
\end{figure}

Figure \ref{dNdA_CMS} shows the differential yield in $A_j$ determined by CMS' dijet sample, compared with \textsc{martini}'s results based on CMS' kinematical cuts \cite{Chatrchyan:2011sx}. 

\section{Conclusions}
\label{conclusions}

The study reported here utilizes 
the pQCD and thermal-QCD based MARTINI numerical simulation with a hydrodynamic background determined by \textsc{music} and 
full jet reconstruction using \textsc{fastjet}.
Using only one free parameter - $\alpha_s$ - we can explain a large part of the jet asymmetries observed
in the recent ATLAS and CMS experiments at the LHC as the consequences of high energy jets interacting with
the evolving QGP medium. The alignment of the dijets $dN/d\phi$ is also well reproduced up to effects resulting from the fluctuating background that we do not yet include.
Further details and computation of the jet fragmentation functions will be presented in a future work.

\section*{Acknowledgments}

CY thanks Jean Barrette, Vasile Topor Pop, and Todd Springer for useful discussions. 
BPS thanks Derek Teaney and Peter Steinberg for helpful discussions.
CG, SJ, and CY were supported by the Natural Sciences 
and Engineering Research Council of Canada. BPS was supported in part by the US Department of Energy under DOE Contract No. 
DEAC02-98CH10886 and by a Lab Directed Research and Development Grant from Brookhaven Science Associates.
%%%%%%%%%%%%%%%%%%%%%%%%%%%%%%%%%%
% \section{Paper Submission}

% Authors should submit their papers to the ePrint arXiv 
% server\footnote{http://arxiv.org/help} 
% after verifying that it is processed correctly by the LaTeX processor.
% Please submit the source code, the style files 
% (revsymb.sty, revtex4.cls, slac\_one.rtx) 
% and any figures; 
% these should be self-contained to generate the paper from source.  

% It is the author's responsibility to ensure that the papers are 
% generated correctly from the source code at the ePrint server. 
% After the paper is accepted by the ePrint server, please verify that
% the layout in the resulting  PDF file conforms to the guidelines 
% described in this document.
% Finally, contact the organizers of DPF-2011 (e-mail: dpf2011@brown.edu ) and your parallel session conveners\footnote{http://www.hep.brown.edu/\texttt{\char`\~}DPF2011/parallel.html} 
% with the ePrint number of the paper; 
% the deadline for submission is 15~September~2011.


\bigskip % extra skip inserted

% \begin{thebibliography}{9}   % Use for  1-9  references
% %\begin{thebibliography}{99} % Use for 10-99 references

% \bibitem{charm07}   http://www.lepp.cornell.edu/charm07/

% \bibitem{templates-ref} http://www.slac.stanford.edu/econf/editors/eprint-template/instructions.html

% \end{thebibliography}

\end{document}